\documentclass[aps,prl,twocolumn,groupedaddress]{revtex4}
\usepackage[section]{placeins}        
\usepackage{float}
 \usepackage{graphics}
 \usepackage{multirow}
\usepackage{epsfig}
\usepackage{pdfpages}
\usepackage{amsmath,amsthm, amssymb, latexsym}
\usepackage{color}
\usepackage{graphics}
\usepackage{epsfig}
\usepackage[outercaption]{sidecap}
\DeclareMathAlphabet{\mathcal}{OMS}{cmsy}{m}{n}
\SetMathAlphabet{\mathcal}{bold}{OMS}{cmsy}{b}{n}

\newcommand{\bee}{\begin{equation}}
\newcommand{\ene}{\end{equation}}
\newcommand{\beea}{\begin{eqnarray}}
\newcommand{\enea}{\end{eqnarray}}

\begin{document}
\title{Excitation of    KdV  Alfven solitons by a  pulsed CO$ _2 $ laser in plasma in the presence of an external magnetic field }
\author{Atul Kumar}
\email{atul.j1211@gmail.com}
\author{Chandrasekar Shukla}
\author{Deepa Verma}
\author{Amita Das} 
\email{amita@ipr.res.in}
\author{Predhiman Kaw$^0$}

\affiliation{Institute for Plasma Research, HBNI, Bhat, Gandhinagar - 382428, India }
\begin{abstract} 
The interaction of laser with plasmas leads to many interesting phenomena which includes laser energy absorption, mode conversion, 
energetic particle generation etc. In this work, the excitation  of Korteweg - de Vries (KdV) Alfven solitons in plasma is  demonstrated with the help of 
2-D Particle - In - Cell  (PIC) simulations. For this purpose,   the propagation of a  pulsed   $CO_2$  (with a wavelength of $10 \mu m$ )
laser normally incident on an overdense 
plasma target  is considered  in the presence of an external magnetic field. The magnitude of the external magnetic field is chosen so as to 
have the electrons magnetized and ions   unmagnetized at the laser frequency.  These solitons propagate stably and 
are observed to be responsible for energetic electron generation as the  background undisturbed electrons of the medium 
get reflected from its front. It is also shown that subsequently, transverse modulations appear in the structure 
which grow with time. 
With recent technological advancements 
in the development of short pulse $CO_2$ lasers  and also on the generation of 
 magnetic fields of the order of tens of kilo - Tesla  in the laboratory, these studies have practical relevance and have a possibility of 
 getting  replicated in laboratory in near future.  

\end{abstract}
\maketitle 

The solitons are nonlinear coherent structures and  are well known for their robustness and 
the characteristic property of preserving their shape  as they propagate  in any medium. 
The generation of soliton in plasma and their role in formation of collisionless shock waves have attracted a significant amount of interest in recent years in a variety of contexts like astrophysics \cite{Langdon,Tidman, Alsop,Chiueh,Begelman}, solar and space physics \cite{Stasiewicz,Sagdeev}, laboratory plasma experiments \cite{Sentoku, Borghesi, Romagnani} etc. They play a key role in charged particle acceleration and plasma heating \cite{1966RvPP....4...23S,Biskamp,Chiueh,Treumann2009,BLANDFORD19871,OHSAWA2014147,Marcowith}. 
In the presence of an external magnetic field,  soliton propagation  with magnetosonic/Alfven  speed in plasma 
have been predicted. They are termed as   magnetosonic/Alfvenic solitons \cite{Gueroult2017}. Such structures have been observed in the Earth's 
magnetosphere by cluster space crafts \cite{PhysRevLett.90.085002} etc. They are also believed to be present in pulsar wind, neutral star and the core 
of active galactic nuclei where the propagation speed is often relativistic \cite{kennel_pellat_1976,stanford1984high}. \footnotetext[0]{Predhiman Kaw is a deceased author.}

A number of studies in the context of short pulse lasers  interacting  with  plasmas  in laboratories  have shown the formation of coherent 
solitonic structures involving electron species only as  in these cases the laser usually  interacts directly 
with the lighter electron species. In  astrophysical context, however,   the ion involvement is dominant. 
The question naturally arises about the possibility of utilizing short pulse lasers (which are considerably handier) for replicating astrophysical phenomena. In this study, we show that it is indeed possible  
and just within technological reach to mimic ion dominated phenomena in laboratory short pulse laser experiments. 
For this purpose, we employ the interaction of long wavelength $CO_2$ lasers with overdense plasmas. An external magnetic field 
is also applied. The magnitude of which is chosen so that while the electrons in the  medium are magnetized, 
the ions remain unmagnetized at the laser frequency. This ensures coupling of the laser energy to ions and its considerable 
role in further dynamics. With the advent of short pulse $CO_2$ lasers and the possibility of magnetic field generation 
of the order of kilo Tesla have already been demonstrated \cite{Sheng_96,Knauer_2010,Yoneda_2012,Nagatomo2013,Fujioka2016,Matsuo2017,Sakata2017,Bailly-Grandvaux2018}, we belive that  this class of experiments 
  to be feasible in near future. 

As of now the  $ CO_2 $ lasers are  efficient industrial lasers in the market for  continuous wave (CW) and relatively long pulse applications  \cite{Beg1997,Haberberger2010,Tochitsky2016}. They have not been as widely explored for the short pulse (sub-nano or pico second)  applications. We demonstrate,  with the help of PIC simulations, here as well as in a series of papers   \cite{Kumar2018}, the immense possibilities  short pulse $CO_2$ lasers offer for physics explorations. In particular, here  we have shown the excitation 
of KdV Alfven soliton and the possibility of accelerating background plasma electrons reflecting from the soliton surface. 

\begin{figure*}[h!]
\center
                \includegraphics[width=\textwidth]{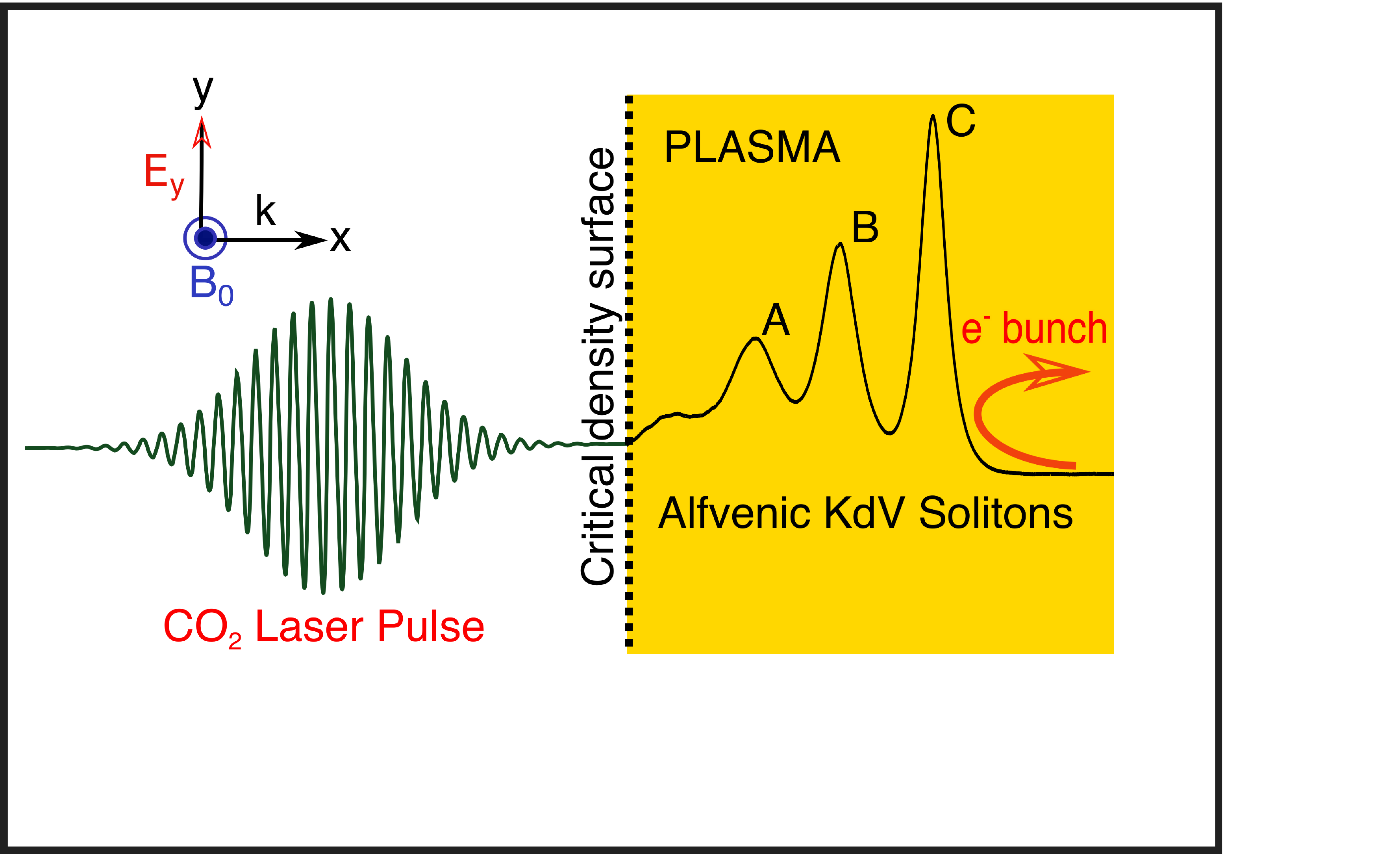} 
             \caption{ Schematic [not to the scale] which shows the system configuration for the excitation of Alfvenic KdV solitons with pulsed $CO_2$ laser in presence of an external magnetic field   }  
                 \label{fig1}
         \end{figure*}

We have carried out PIC simulations under OSIRIS-4.0 framework \cite{hemker,Fonseca2002,osiris}.  
 We have considered a  system  of electrons and ion plasma  and chosen for simplicity and faster simulations,
  ions  to be $25$ times heavier than electrons (\emph{i.e.} $m_i = 25 m_e$, where $m_i$ and $m_e$ denotes the rest mass of the ion and electron species). For electrons to be strongly magnetized and ions to remain unmagnetized at the laser frequency, we have considered  the magnetic field strength in such a fashion so as to satisfy the condition of $ \omega_{ci} < \omega < \omega_{ce}$, where $\omega$ is the laser frequency. 
In particular we have chosen 
$    \omega_{ci} = \omega/2 $. The mass ratio being $25$ this implies  $\omega_{ce} = 12.5 \omega$ which ensures that 
the  electrons  are 
strongly magnetized. A rectangular geometry of  $ 1500 c/\omega_{pe} \times 100 c/\omega_{pe} $   in $ x-y $ plane has been choosen for the simulation where plasma boundary starts from $500 c/\omega_{pe}$. Thus there is  vacuum between $ 0 $ and $500 c/\omega_{pe}$. The plasma density is chosen such that 
the laser frequency $\omega = \omega_{pi}$ resonates with the ion plasma frequency. 
The plasma density is chosen to be at the critical ion density 
and $ \omega_{ce} = 2.5 \omega_{pe} $. 
The spatial resolution chosen in the simulation is $10$ cells per electron skin depth with $64$ particles per cell for each species corresponding to a grid size $\Delta x = 0.1 c/\omega_{pe}$ and time step $ \Delta t = 0.07 \omega_{pe}^{-1} $. The boundary conditions for electromagnetic fields and particles are  periodic in $\hat{y}$  direction and absorbing in $\hat{x}$ direction. A p-polarized, plane X-mode $ CO_2 $ short pulse laser of wavelength $10 \mu m$,  is incident from the left boundary at a sharp plasma surface at critical density along $ \hat{x} $ direction. The laser peak intensity taken in the simulation is $I = 7\times 10^{17} W/cm^2$ with a rise and fall time of $204.6217 \omega_{pe}^{-1}$  each. A  uniform and static external magnetic field of $B_{0} =14.13$ kilo-Tesla in transverse direction ($\hat{z}$ direction) has been applied in the system.  The system has been illustrated in the  schematics of  Fig.~({\ref{fig1}}).  
 
\begin{figure*}[h!]
\center
             \includegraphics[width=\textwidth]{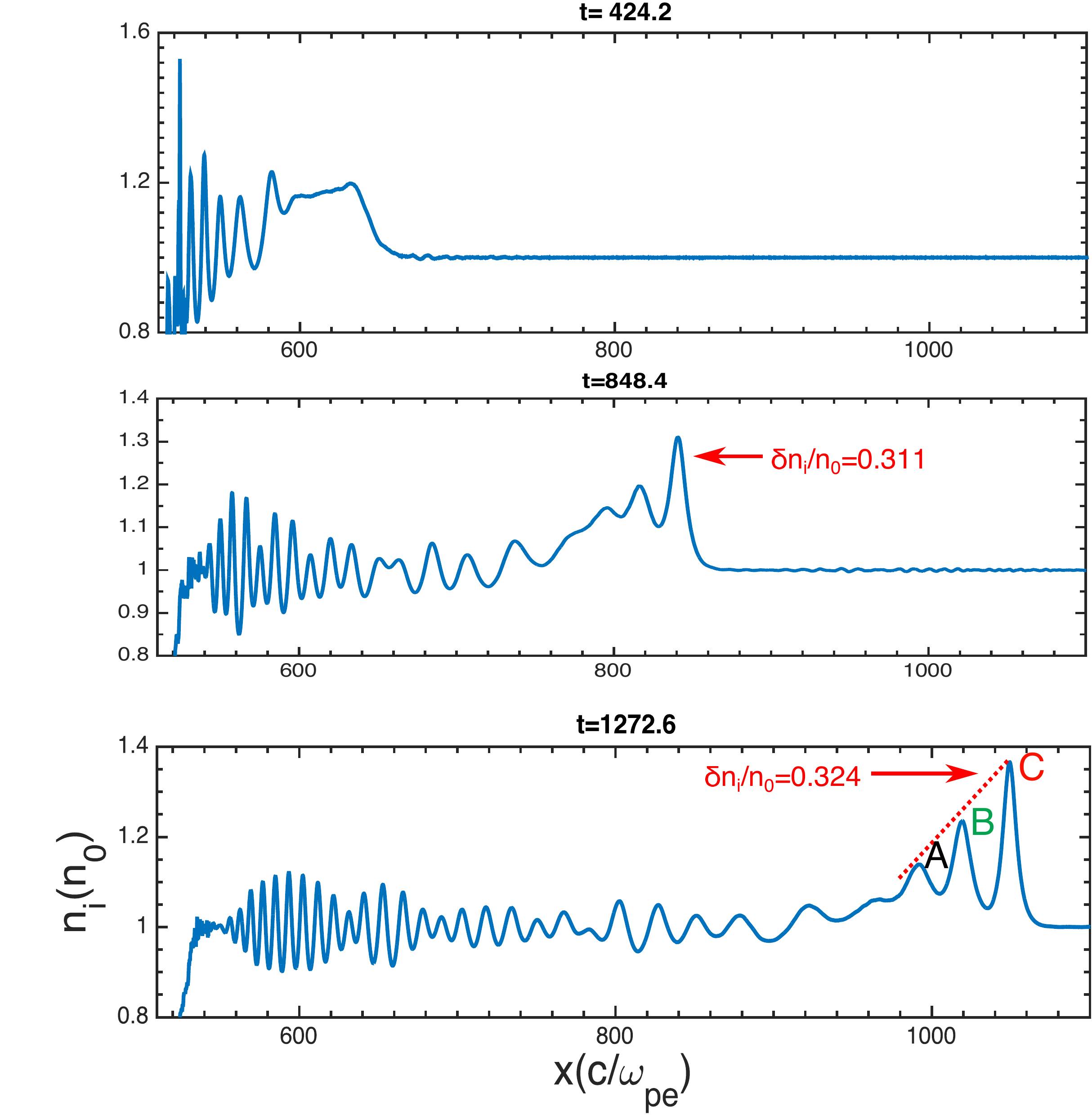} \\   
             \caption{Time (in $\omega_{pe}^{-1}$) evolution of  ion density where it is shown that three coherent structures $A$, $B$, and $C$ are formed at $t=1272.6 \omega_{pe}^{-1}$ }  
                 \label{fig2}
         \end{figure*}

 A p -polarised laser incident normally over the sharp plasma interface in an unmagnetized plasma would not be able to dump its energy to  the plasma. However, in the presence of external magnetic field under the condition of 
 $\omega_{ci} < \omega < \omega_{ce}$ the ponderomotive pressure of the laser pulse acts along different direction for the ions and electrons \cite{Nishikawa}. This leads to a charge separation. This charge separation triggers ion plasma oscillations. Since the  
  laser frequency is tuned to ion plasma period these oscillations are resonantly excited. The electrons being tied to the magnetic field 
  are, however, unable to respond to this electrostatic field. These electrostatic disturbances are clearly evident in Fig.~(\ref{fig2}) which shows the 
  plot of temporal evolution of ion density in $\hat{x}$ direction.  
  It can also be observed from Fig.~(\ref{fig2}) that the high amplitude laser pulse triggers a large amplitude plasma  disturbance. 
  This disturbance is seen to break in three pulses and propagate reasonably stable inside the plasma target. This coherent structure formation and its stable propagation is observed in all the fields, namely the  ion density ($n_i$), electron density ($n_e$), transverse electric field $E_y$ (transverse electric field) and magnetic field $B_z$. 
  The three pulses  termed as  ($A$, $B$ and $C$ ),    in all these fields are observed to arrange in increasing amplitude 
  with time in a straight line. This happens as the speed of the soliton is proportional to the amplitude 
  and hence the  higher amplitude soliton moves faster and overtakes the low amplitude solitonic structures. 
%
  We measured the propagation speed of these solitonic structures from our simulations 
  and found it to be $v_{sim, A}=0.4620c$,$v_{sim,B}=0.4787c$,  and $v_{sim,C}=0.4933c$ for the three structures $A$, $B$, and $C$ respectively.  These propagation speed match closely with  the Alfven speed  of the medium which 
  is  $v_A=(m_e/m_i)^{-1/2}\omega_{ce}/\omega_{pe}= 0.5c$ for our choice of parameters of the medium.
 The KdV solitons \cite{Zubasky} also have an interesting property that  for different solitons, the amplitude $a$ and the width $L$ of the solitons vary in such a fashion so as to have 
 $aL^2$ as constant. We have evaluated the same for the three structures observed in the simulation at various time 
 which is shown in TABLE I.


\begin{center}
{\bf{TABLE I}} \\

The parameter $ aL^2 $ remains constant for MS solitons $A$, $B$ and $C$ with time\\
\vspace{0.2in}
\begin{tabular}{c c c c c c c c c c c c c  ll}
\hline
\hline
  &$Soliton $ \hspace{0.15in}  &$t (\omega_{pe}^{-1}) $ \hspace{0.15in}   & $a (\delta n_i/n_0)$  \hspace{0.15in}        &$L(c/\omega_{pe})$ \hspace{0.15in}        &$aL^2$  \\   
 \hline
  &  \hspace{0.15in}      & 1272.6  \hspace{0.15in}    &   0.065   \hspace{0.15in}      &16.0   \hspace{0.15in}    &   16.64 \\
  & A    \hspace{0.15in}      & 1484.7 \hspace{0.15in}     & 0.072  \hspace{0.15in}      &16.0 \hspace{0.15in}    &   18.43 \\
  &     \hspace{0.15in}      & 1696.8  \hspace{0.15in}    & 0.081  \hspace{0.15in}    &15.0  \hspace{0.15in}    &   18.22 \\
  \hline

&  \hspace{0.15in}      & 1272.6  \hspace{0.15in}    &   0.16   \hspace{0.15in}      &10.0   \hspace{0.15in}    &   16.00    \\
  & B    \hspace{0.15in}      & 1484.7 \hspace{0.15in}     & 0.18 \hspace{0.15in}      &10.0 \hspace{0.15in}    &   18.10   \\
  &     \hspace{0.15in}      & 1696.8  \hspace{0.15in}    & 0.194  \hspace{0.15in}    &9.6  \hspace{0.15in}    &   17.88    \\
  \hline
  
   &  \hspace{0.15in}      & 1272.6  \hspace{0.15in}    &   0.324   \hspace{0.15in}      &7.5   \hspace{0.15in}    &   18.22 \\
  & C    \hspace{0.15in}      & 1484.7 \hspace{0.15in}     & 0.336  \hspace{0.15in}      &7.6 \hspace{0.15in}    &   19.40    \\
  &     \hspace{0.15in}      & 1696.8  \hspace{0.15in}    & 0.337  \hspace{0.15in}    &7.6  \hspace{0.15in}    &  19.46     \\
\hline
\end{tabular}
  \end{center} 
It is clear from the TABLE I that despite  the  difference in the amplitudes  and width of each of the solitons,  the parameter $aL^2$ remains relatively unaltered. While the amplitude $a$ of the different soliton differ by as much as the value of $ a $ is about 
$ 81\% $, and the square of the width  $ L^2 $ by  $ 77\% $; the variation in $aL^2$ in the observations is  only about $14.5 \%$. 
This clearly suggests that the structures which have formed in the simulations are essentially KdV solitons. The slight variations 
occur as the conditions of the medium changes with time and the KdV solitons are the solutions in the 
reductive perturbative expansion of the full set of MHD equations in the weakly dispersive, nonlinear regime. These studies, therefore, confirm that the Alfvenic KdV solitons can be excited in a laboratory by using pulsed $CO_2$ lasers along with an externally applied magnetic field. 

We have also observed that as the large amplitude soliton moves through the plasma, the background electrons colliding with it acquire 
energy.  Electron bunches are created as shown in  Fig.~(\ref{fig5}). In fact the acceleration process 
has been found to be pretty efficient with these electrons acquiring 
a very high relativistic speed of $\approx 0.94 c$ and moving ahead of the solitonic structures. The electrostatic field associated with electron bunches have been shown in 
 Fig.(\ref{fig6}). Fig.~(\ref{fig6}) also confirms the generation of electron ahead of the Alfvenic KdV solitons.  The electron bunch is a highly energetic about $ 10 MeV$ and collimated as shown in Fig.~(\ref{fig7}). Ions being very massive than electrons, do not get excited by the front of the soliton and thus remain stationary shown in Fig.~(\ref{fig8}).
 
It is well known that in higher dimensions the soliton formation with cubic nonlinearity of KdV equation is not permitted. 
The structures instead show a tendency to collapse. 
The PIC simulations being carried out in 2-D, at later stages shows a development of transverse modulations 
on soliton surface. These modulations are observed to  increase with time 
  as shown in Fig.~(\ref{fig9}), signifying a development of an instability leading towards the collapsed state in 2-D. However, 
  the complete physics of the real system has components which are  beyond KdV and hence  the transverse  collapse of the system 
  is captured.  
  
  To conclude, we have demonstrated the possibility of capturing  ion dynamics dominated phenomena in short pulse lasers by 
  employing the low frequency $CO_2$ lasers along with applied magnetic fields. In this work with the help of PIC simulations 
  using OSIRIS4.0 we have shown the excitation of coherent structures involving ions, which has been 
   identified as the  Alfvenic KdV solitons.  These Alfvenic solitons are observed to propagate stably for several ion plasma 
    periods ($\sim382\omega_{pi}^{-1}$). At later times ( $\sim t=1696.8 \omega_{pe}^{-1}$), transverse modulations in the structures appear which grows with time. 
    However, this modulations do not lead to a complete collapse of the  structures. 
   Our simulations have also shown the possibility of employing these structures for electron acceleration.  
 
\bibliographystyle{ieeetr}  

\bibliography{magnetosonic}

          \begin{figure*}[h!]
\center
                \includegraphics[width=\textwidth]{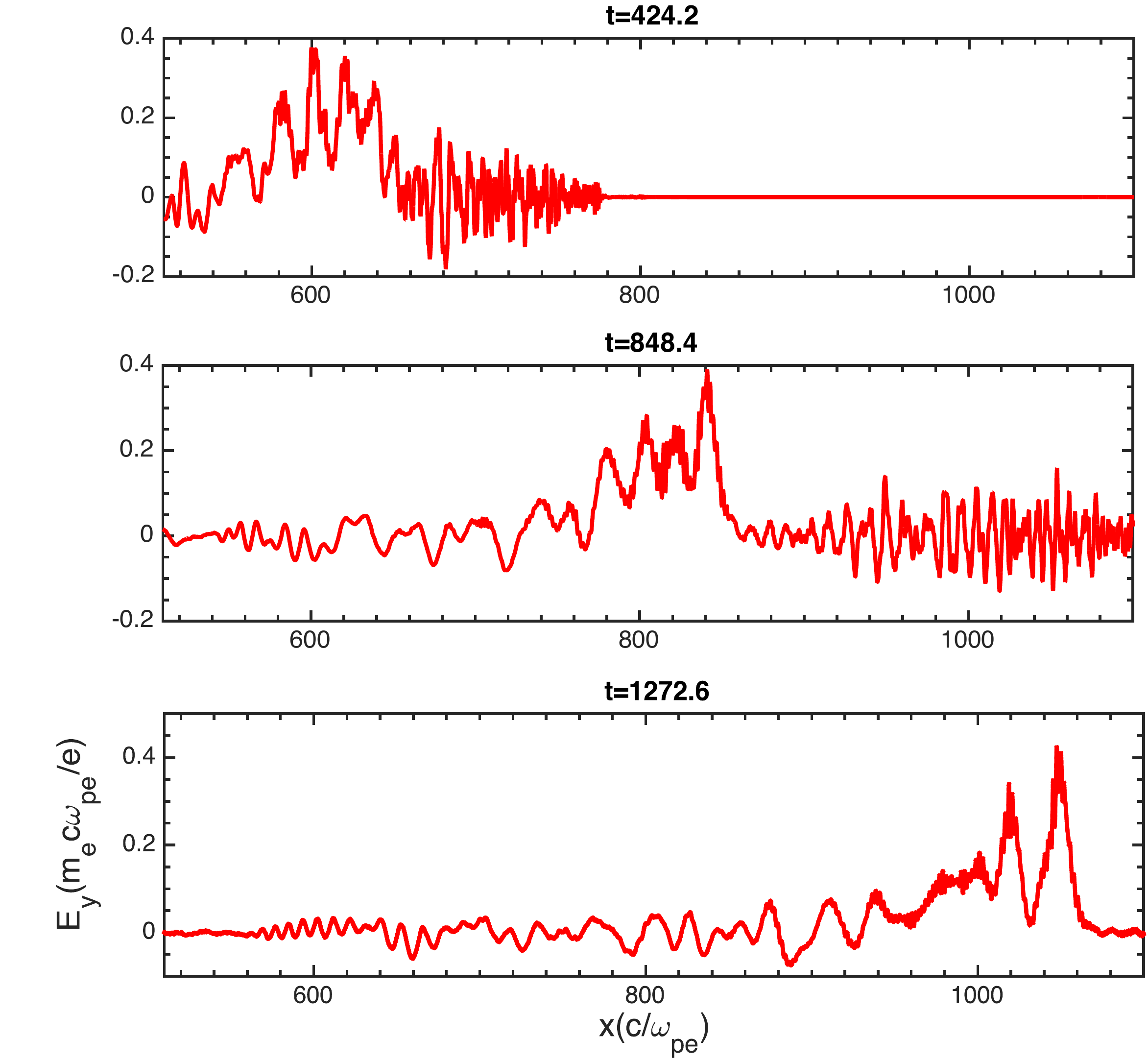} 
                
             \caption{ Time (in $\omega_{pe}^{-1}$) evolution of the electromagnetic component of electric field which is generated due to Alfvenic KdV solitary wave}

                 \label{fig3}
         \end{figure*}   
           \begin{figure*}[h!]
\center
                \includegraphics[width=\textwidth]{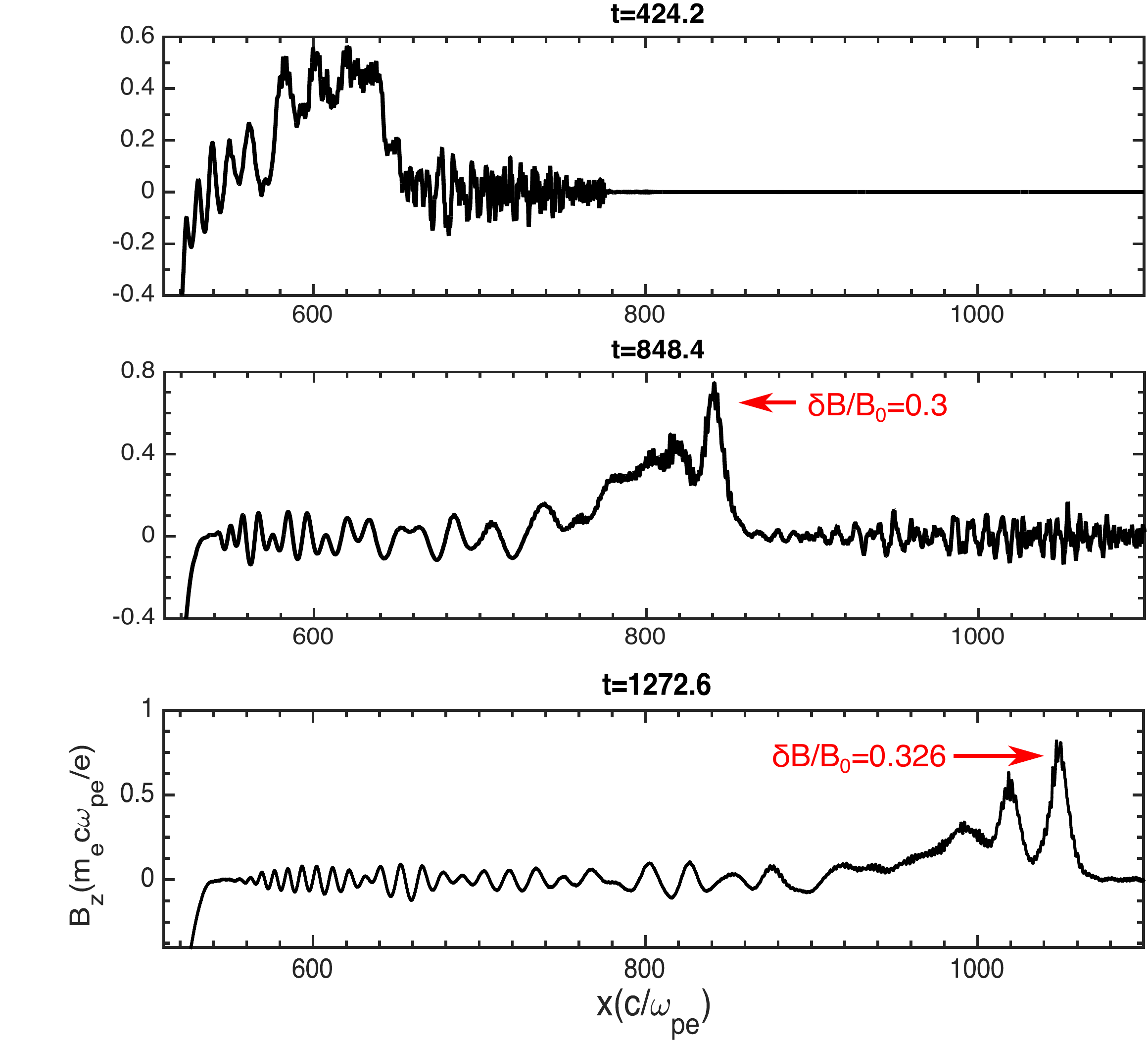} 
             \caption{Temporal evolution of the perturbed magnetic field $B_z$ corresponding to the Alfvenic KdV solitons}  
                 \label{fig4}
         \end{figure*} 
         
          \begin{figure*}[h!]
\center
                \includegraphics[width=\textwidth]{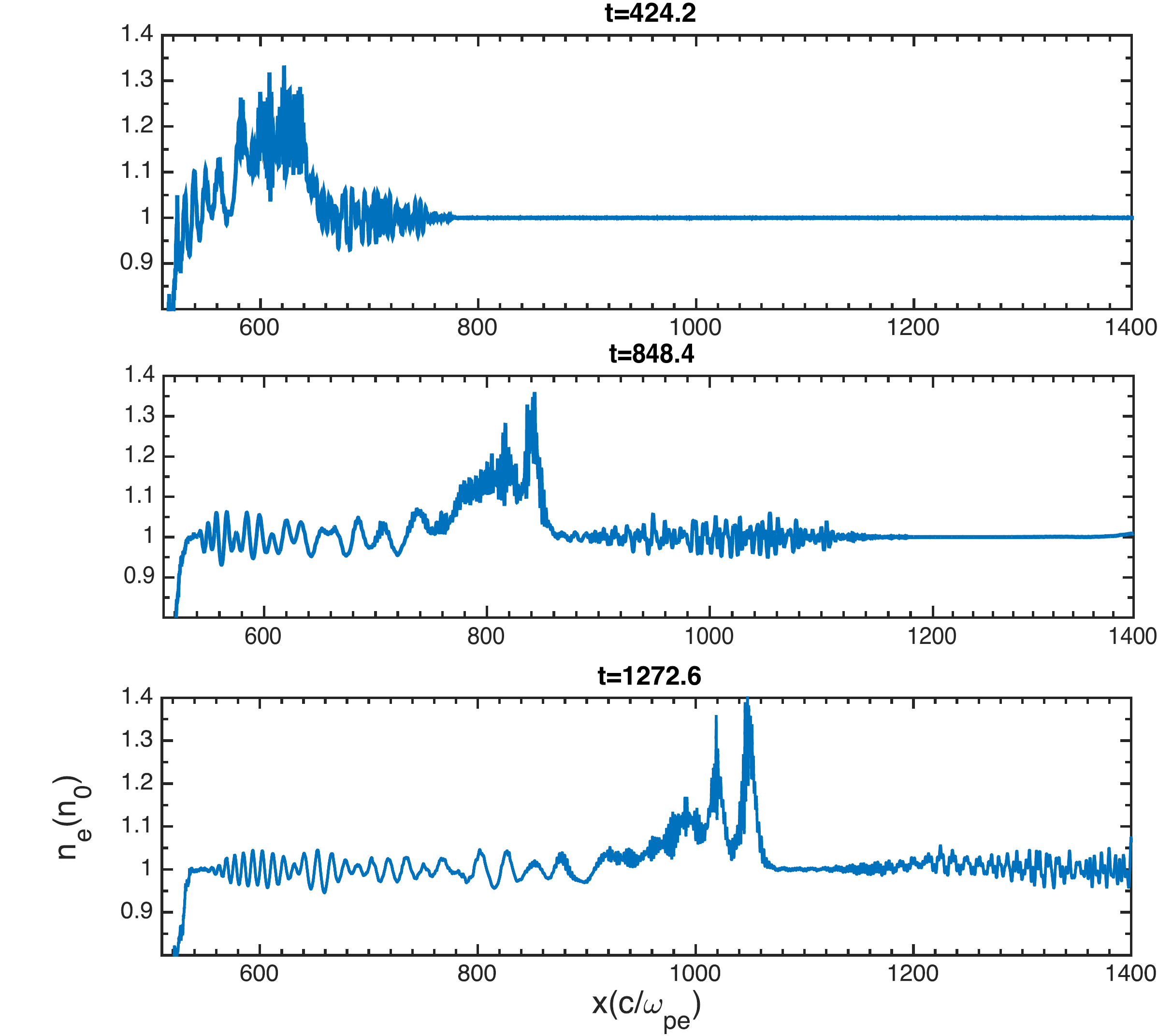} 
             \caption{Temporal evolution of the electron charge density. The coherent structures can also be seen in the electron charge density  }

                 \label{fig5}
         \end{figure*} 
            \begin{figure*}[h!]
\center
                \includegraphics[width=\textwidth]{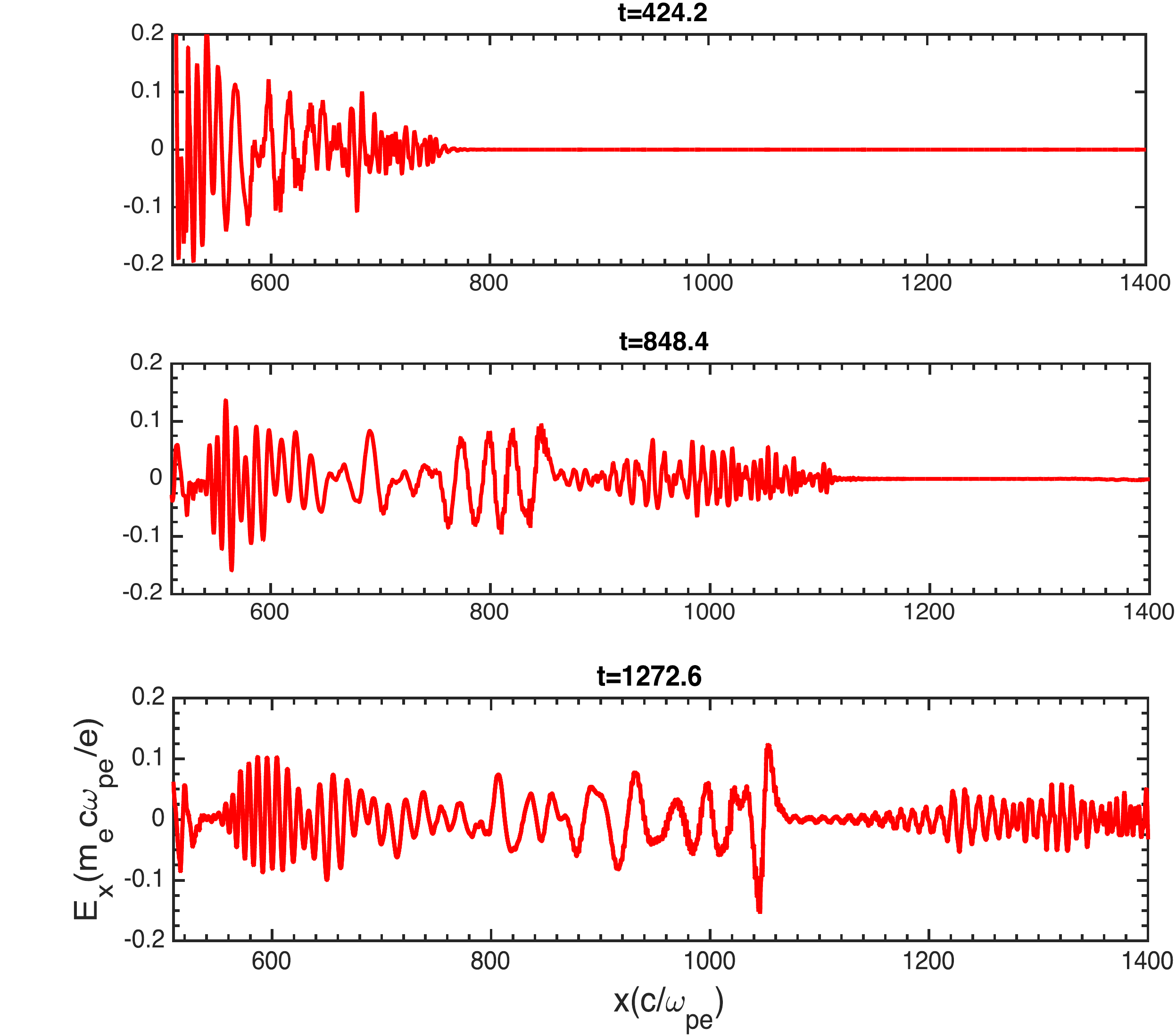} 
             \caption{Time evolution of the electrostatic component of the electric field. This  electric field generated at and/or beyond the front of solitons are due to the cold electrons that get reflected by the large solitary pulse}  
                 \label{fig6}
         \end{figure*}

         \begin{figure*}[h!]
\center
                \includegraphics[width=\textwidth]{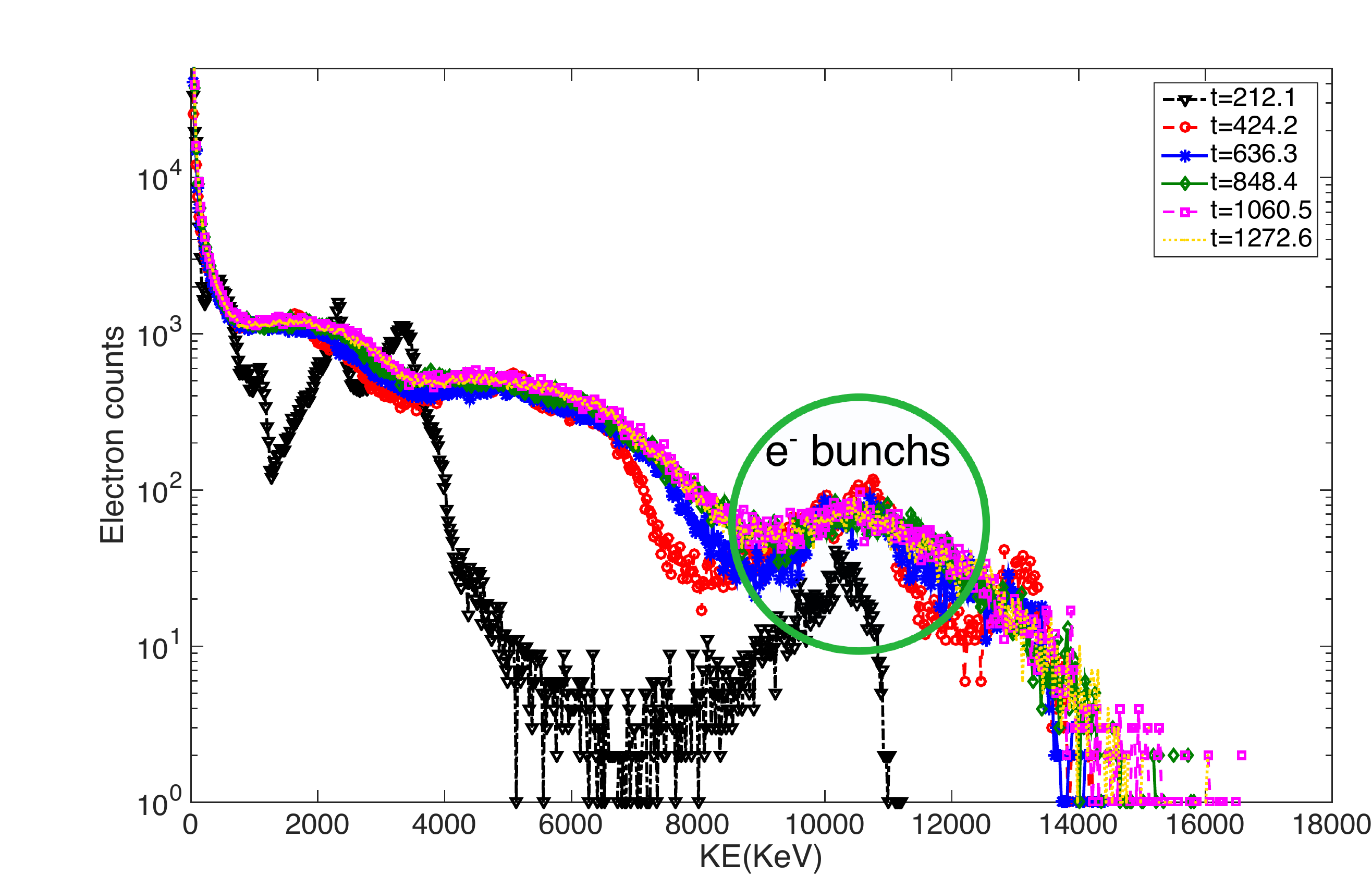} 
             \caption{ Energy distribution of the electrons which shows that the quasi-monoenergetic electron bunch are generated by the Alfvenic KdV solitons }  
                 \label{fig7}
         \end{figure*}
         
          \begin{figure*}[h!]
\center
                \includegraphics[width=\textwidth]{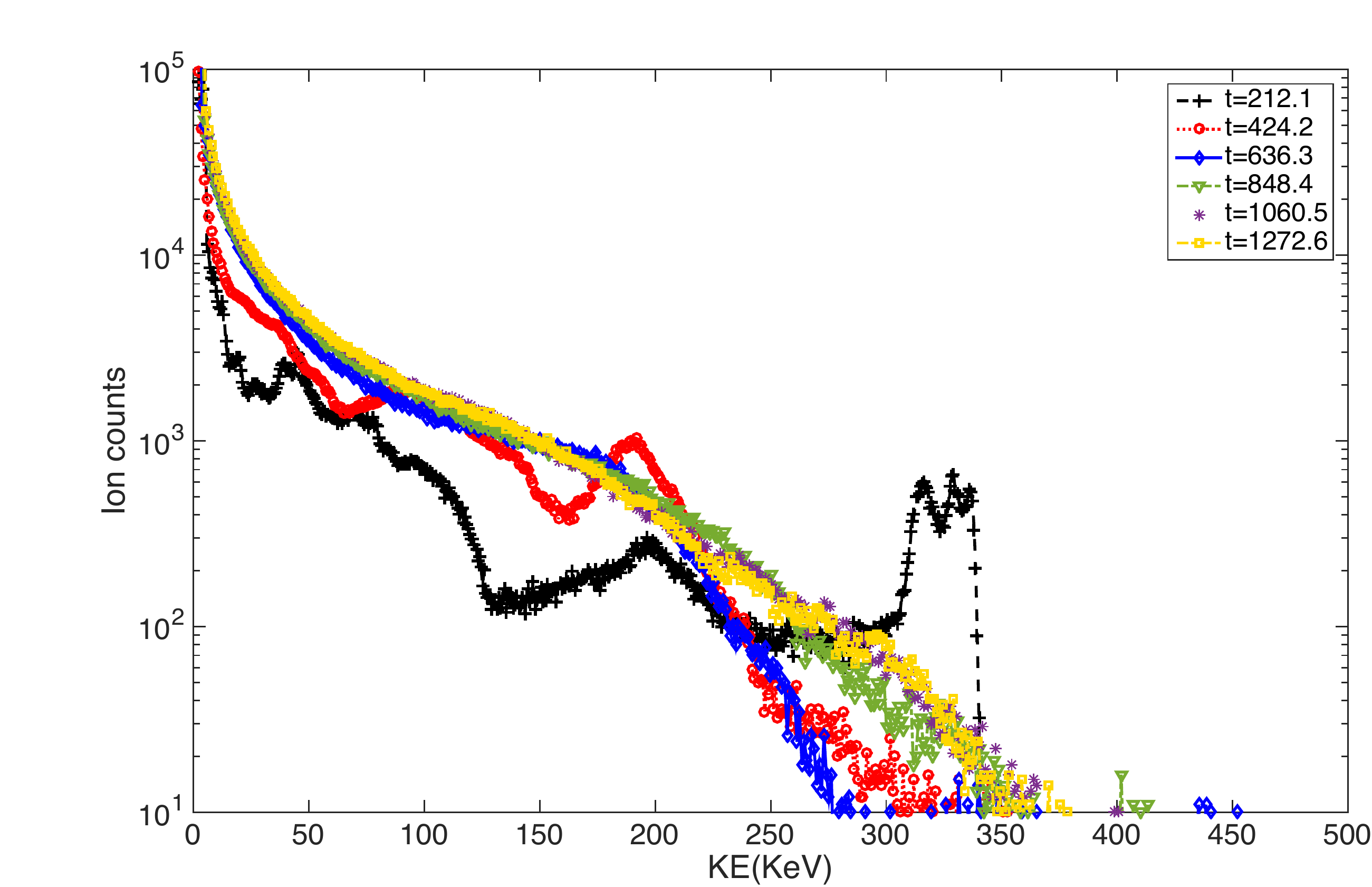} 
                \caption{ Energy distribution of ions which clearly shows that ions remain static and do not acquire much energy from the front of the solitons  }  
             
                 \label{fig8}
         \end{figure*}
        \begin{figure*}[h!]
\center
                \includegraphics[width=\textwidth]{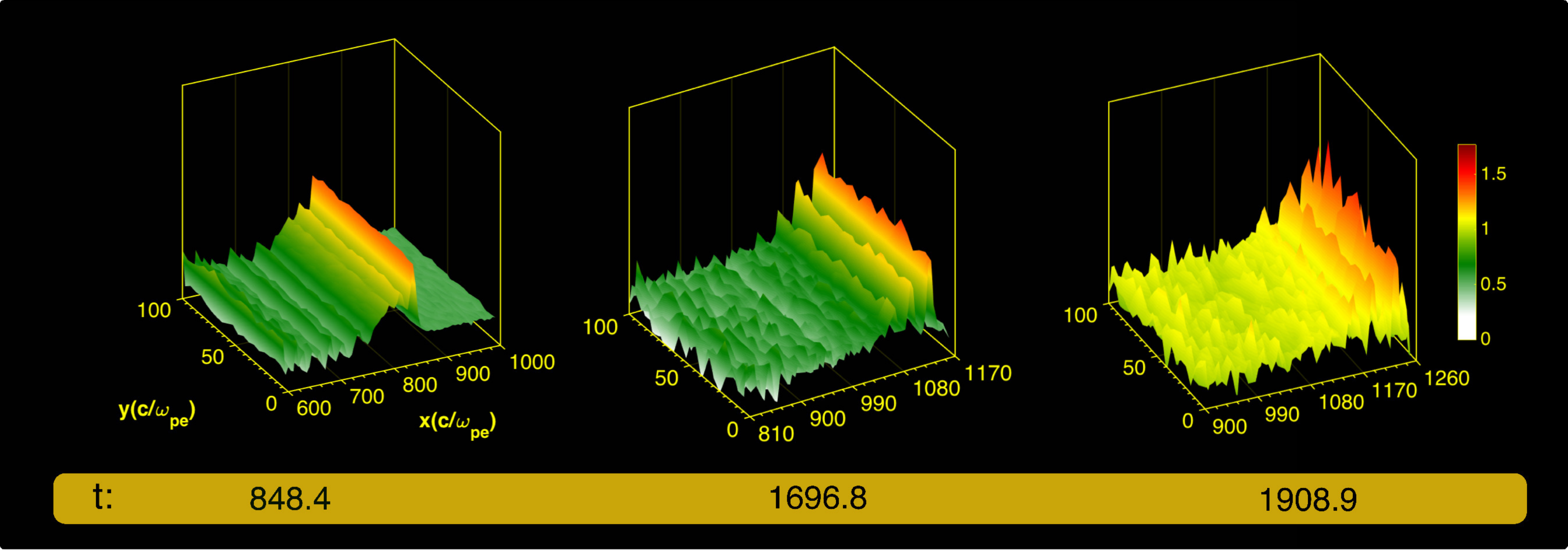} 
                \caption{ Transverse filamentation of each peak of the  ion charge density of the Alfvenic KdV soliton; It has been shown that at $t=1696.8 \omega	_{pe}^{-1}$, transverse filamentation  starts and it continues to grow with time  }  
             
                 \label{fig9}
         \end{figure*}
         
%
%
%
%

\end{document}